\input{aipcheck}

\documentclass[
    ,final            
]
  {aipproc}

\layoutstyle{8x11double}

\begin{document}

\title{On the Ordering of Na$^+$ Ions in Na$_x$CoO$_2$}

\classification{
                \texttt 71.27.+a; 71.45.Lr}

\author{M. Roger}{
  address={DSM DRECAM Service de Physique de l'Etat Condens\'e,
CEA Saclay, 9191 Gif sur Yvette, France}
}

\author{D.J.P. Morris}{
address={Department of Physics, University of Liverpool, Oliver Lodge Laboratory
Liverpool L69 7ZE, UK}
}

\author{D.A. Tennant}{
  address={School of Physics and Astronomy, North Haugh, St Andrews,
Fife KY16 9SS UK}
}

\author{M.J. Gutmann}{
address={ISIS Facility, Rutherford, Appleton Laboratory, Chilton,
Didcot, Oxon, OX11 0QX, UK}
}

\author{J.P. Goff}{
address={Department of Physics, University of Liverpool, Oliver Lodge Laboratory
Liverpool L69 7ZE, UK}
}

\author{D.~Prabhakaran}{
address={Clarendon Laboratory, Parks Road, Oxford OX1 3PU, UK}
}

\author{N. Shannon}{
address={Univ. Tokyo, Dep. Adv. Mat. Sci., Grad. Sch. Frontier Sci.,
5-1-5 Kashiwahnoha, Chiba 2778851 Japan}
}

\author{B. Lake}{
address={Clarendon Laboratory, Parks Road, Oxford OX1 3PU, UK}
}

\author{A.T. Boothroyd}{
address={Clarendon Laboratory, Parks Road, Oxford OX1 3PU, UK}
}

\author{R. Coldea}{
address={Clarendon Laboratory, Parks Road, Oxford OX1 3PU, UK}
}

\author{P. Deen}{
address={European Synchrotron Radiation Facility, BP 220, 38043 Grenoble
Cedex, France}
}



\begin{abstract}
The influence of electrostatic interactions on the ordering of sodium 
ions in Na$_x$CoO$_2$ is studied theoretically through Monte-Carlo 
simulations. For large $x$  small di- or tri-vacancy clusters are stable
with respect to isolated Na vacancies. At commensurate fillings these
small clusters order in triangular superstructures. These results 
 agree with recent electron diffraction data at $x=1/2$ and
$3/4$. We have performed neutron La\"ue diffraction experiments
at higher $x$, which confirm the predictions of this simple model.
The consequences on the properties of  the electronic charges 
in the Co layers are discussed.  
\end{abstract}

\maketitle


\begin{figure}[b]
  \includegraphics[width=.17\textheight]{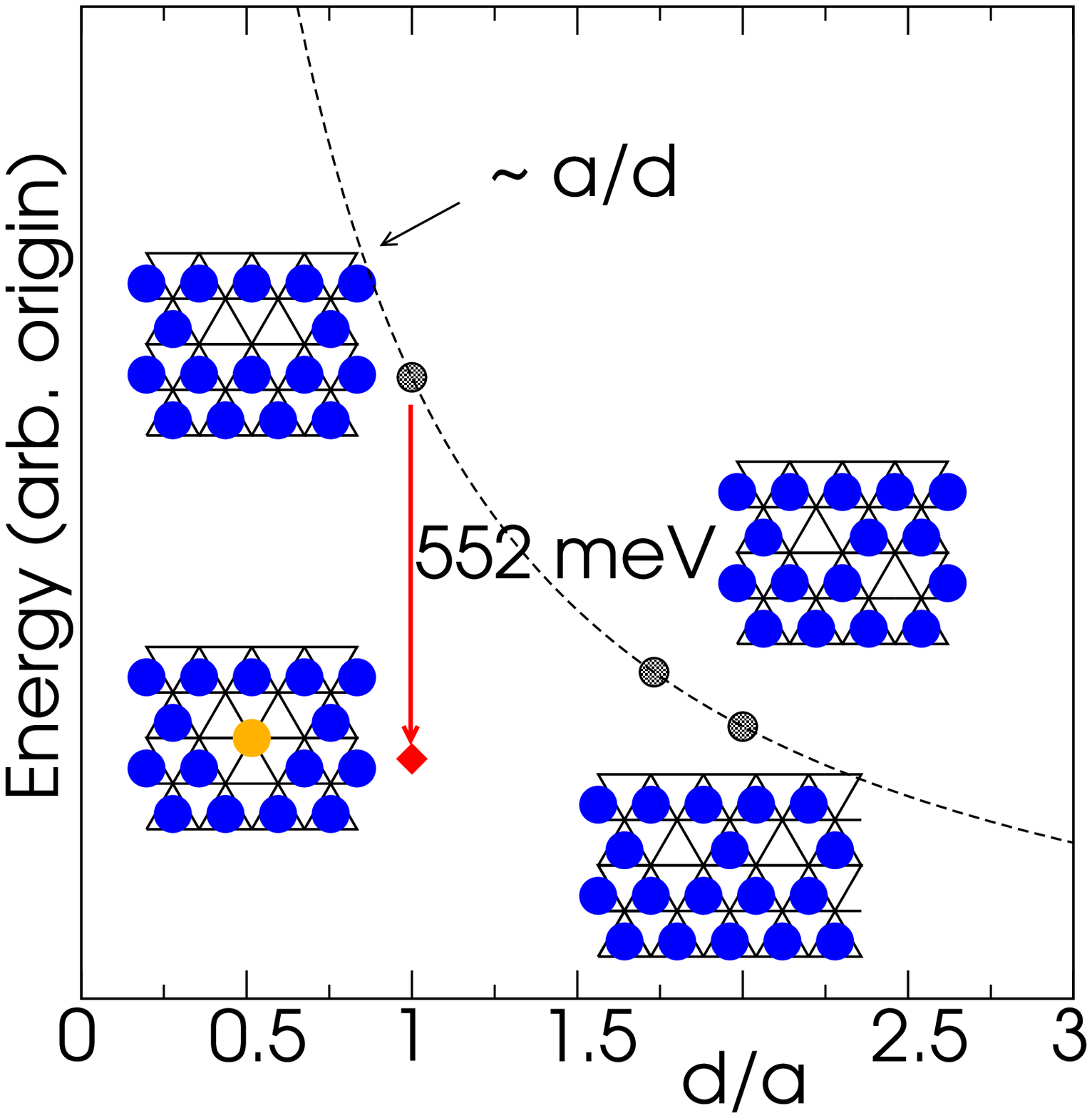}
  \includegraphics[width=.17\textheight]{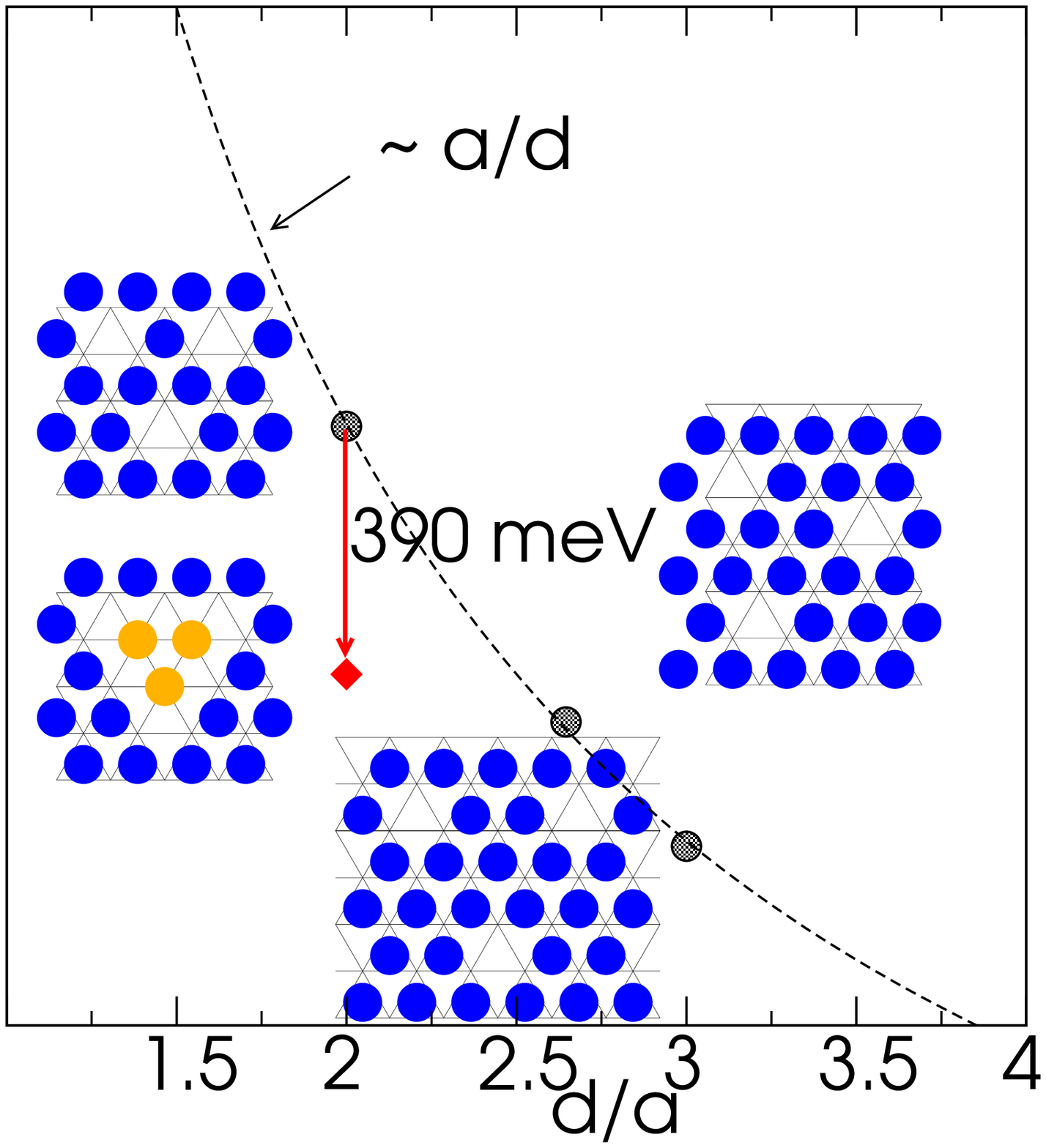}
  \caption{Di- and tri-vacancies. Dark (light) circles correspond
to occupied Na$_2$ (Na$_1$) sites}
\end{figure}

In the past decade, Na$_x$CoO$_2$ has emerged as a system of fundamental
interest because of its high thermoelectric power, unusual magnetic
properties and possibility, when hydrated, to superconduct \cite{Takada}.
 The density $x$
of intercalated sodium atoms tunes directly the density of states of 
quasi-particles in the metallic CoO$_2$ sheets. For simple fractional
f\i llings, electron diffraction (ED) experiments \cite{Zandbergen,Shi}
provide evidence for the ordering of Na$^+$ ions at room temperature. This
non-trivial order is generally believed to be the consequence of intricate
interactions between Na$^+$ ions, mobile electrons, and phonons. We show here
that the dominant process driving this ordering is simply the 
Coulomb potential between Na$^+$ ions. Sodium ions can occupy sites  
between oxygen atoms, which form two interpenetrating triangular lattices
denoted Na$_1$ and Na$_2$. Cobalt ions lie above and below Na$_1$ sites, 
resulting in an extra energy cost $\Delta=\Delta_c+\Delta_{sr}$ relative
to an Na$_2$ site ($\Delta_c$ in the long-range Coulomb part and 
$\Delta_{sr}$ a short range repulsion between Na and Co ions shells).
Na ions are bigger than the distance between nearest Na$_1$ and Na$_2$
sites, forbidding simultaneous occupancy of nearest-neighboring sites and
that makes Coulomb energy minimization highly non-trivial. Our model
Hamiltonian includes: (i) a long-range Coulomb potential, 
$ e/(4\pi\epsilon\epsilon_0 r) $ (the dielectric constant, taken as isotropic,
is f\i xed at $\epsilon=6$ to account quantitatively for the variations of
the chemical potential, in terms of $x$, measured in cell devices \cite{Amatucci}); 
(ii) Na-Na ion-shell repulsion V=0.04eV for neighbors on the same sublattice;
and (iii) on-site energy $\Delta_{sr}=0.01$eV. A system of two sheets of Na ions,
each containing a maximum number of 1176 ions, intercalated between two CoO$_2$
layers (with periodic boundary conditions in 3 dimensions) is simulated at 
f\i nite temperature and f\i xed chemical potential $\mu$ through
a Grand-Canonical Monte-Carlo method. The hexagonal cell parameters $a$=2.84\AA ~
and $c$=10.87\AA ~ are f\i xed to those measured in the $x=0.75$ phase, and elastic
deformations are neglected. We maintain neutrality at each step by varying the
charges in the Co layers assuming a uniform spreading.

The organization of Na ions is driven by the spontaneous formation of
multivacancy clusters, as illustrated in Figure 1. At large distance $d$, the 
repulsion energy between two vacancies decreases as $~1/d$. However
neighboring vacancies can reduce their energy by promotion of a Na$_2$ sodium
to the central Na$_1$ site. The resultant cluster has a net charge 2e$^-$ spread
over three sites and substantially lower energy, as the central sodium is now
further from its neighbors. Formation of a tri-vacancy cluster follows a
similar process and three vacancies combine with three Na$_2\to$ Na$_1$ promotions.

Monte-Carlo results at room temperature are presented in Figure 2. The Na 
concentration versus chemical potential is characterized by plateaus
at $x={1\over3},~ {1\over 2},~{ 5\over 7},~{ 3\over 4}$, comparable to those 
appearing in electrochemical cells \cite{Amatucci}.
The the ${1\over 2}$-plateau corresponds 
to an \{{\bf a'=2a-b,~b'=2b}\} ordered structure with
equal occupation of Na$_1$ and Na$_2$ sites in perfect agreement with ED
spectra \cite{Zandbergen}. 
It can be described as a dense lattice of divacancy clusters.
The ${5\over 7}$-structure is 
a  \{{\bf a'=2a+b,~b'=-a+3b}\} triangular array of divacancies.
At $x={3\over4}~$  a triangular array  \{{\bf a'~=~~2a~+~2b, \\ b'=~-2a+4b}\}
of trivacancies is obtained, in agreement with ED \cite{Zandbergen,Shi}.  
No plateau is observed at higher $x$, but short-range ordered random mixing of
di- trivacancies appears in the simulations. Nevertheless, at $x>0.8$ many
commensurate triangular arrays of di- trivacancies 
are possible and we have therefore
investigated experimentally this range though neutron diffraction.

\begin{figure}[t]
  \includegraphics[width=.35\textheight]{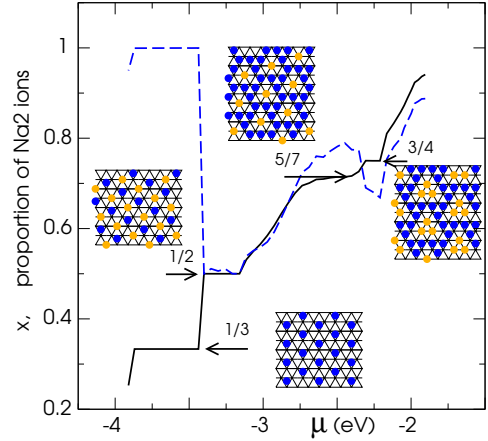}
  \caption{Na concentration $x$ (solid line) and relative proportion
of Na$_2$ sites (dashed line) versus chemical potential $\mu$}
\end{figure}

\begin{figure}[b]
  \includegraphics[width=.1\textheight]{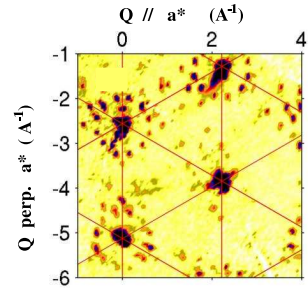}
  \caption{Neutron spectrum in (h,k,6) plane at $x\approx0.92$}
\end{figure}

Single crystal data were collected on the SXD diffractometer at the ISIS
pulsed neutron source, Rutherford Appleton Laboratory (UK) 
with area detectors covering a solid angle of $2\pi$
steradians. At $x=0.92\pm0.03$, Figure 3 shows superstructure
peaks appearing on a circle of radius $\tilde q \approx 6.16 nm^{-1}$ centered
on the main Bragg points. These are compatible with a triangular array of
divacancies with  unit-cell vectors \{{\bf a'=3a+b,~b'=-a+4b}\}. 

Sodium ordering induces a periodic Coulomb potential in the Co layers. The
depth of the electrostatic potential wells, $\approx$100meV, is substantially larger
than the quasi-particle hopping frequency $\approx$10meV and so 
will localize holes.
While total energy differences are insignif\i cant ($\approx$0.1meV), the Coulomb
landscape varies drastically in terms of position of vacancy clusters in
successive layers along c. For $x={1\over 2}$ there are only two conf\i gurations.
Figure 4~(a) corresponds to divacancies in successive layers displaced
by {\bf b}. It has a one-dimensional character with high (dark) and low (light)
 Coulomb-potential
stripes in agreement with the low and high spin charge-ordered stripes found
using neutron diffraction \cite{Yokoi}. A large magnetic f\i eld parallel to 
the ({\bf a,b})
plane favors alignment of holes (divacancies) in the {\bf c}-direction corresponding
to Figure 4~(b) with 2D character and conducting Co ions at intermediate valencies
(gray circles). This could explain the abrupt
change from an insulating antiferromagnet to a conducting state with 
reconstruction of a 2D Fermi surface when a f\i eld of 40T is 
applied at T=50K \cite{Balicas}. The potentials corresponding to the $x={5\over 7}$
 and $3\over 4$ structures are discussed in Ref. \cite{Alan}.
 
In conclusion, at $x>{1\over 2}$,
 the ordering of sodium ions governs the electronic and
magnetic properties of quasi-particles (holes) in the Co layers. Pure electrostatics
with non-trivial constraints on a bipartite lattice leads to vacancy clustering.
For simple fractional f\i llings di- and trivacancy clusters order long range.

\begin{figure}[t]
  \includegraphics[width=.1\textheight]{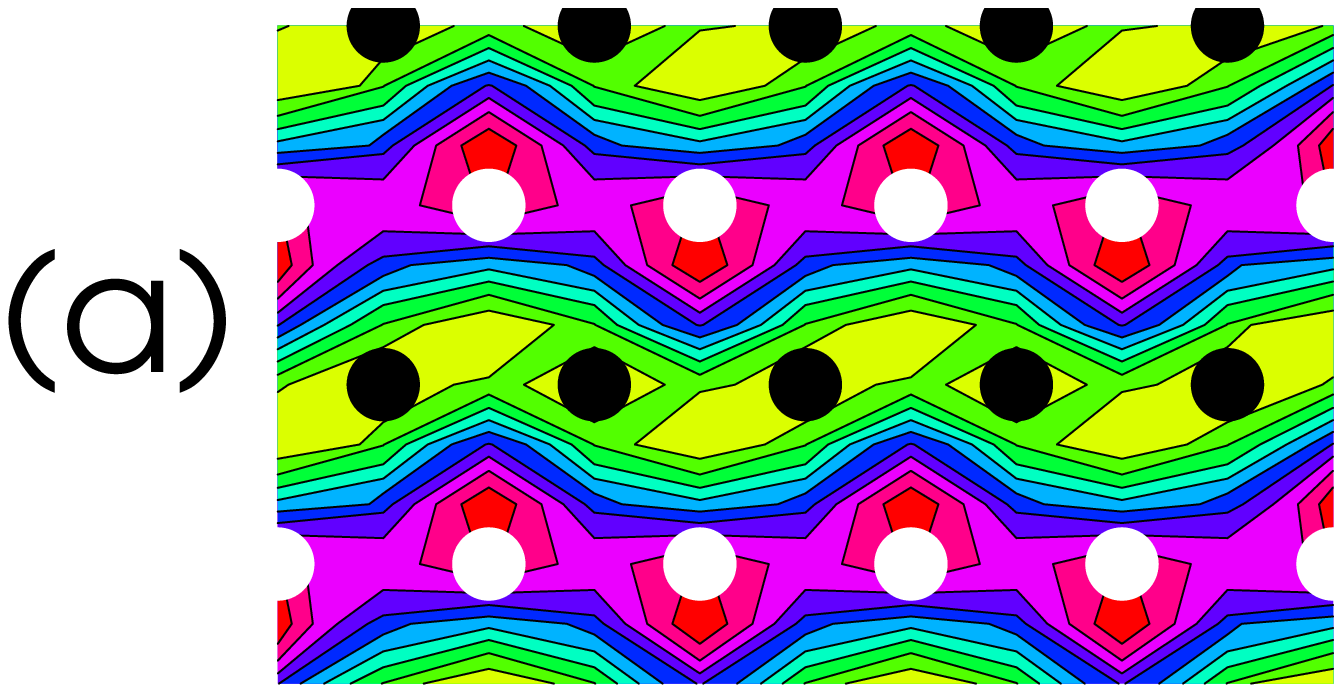}
  \includegraphics[width=.1\textheight]{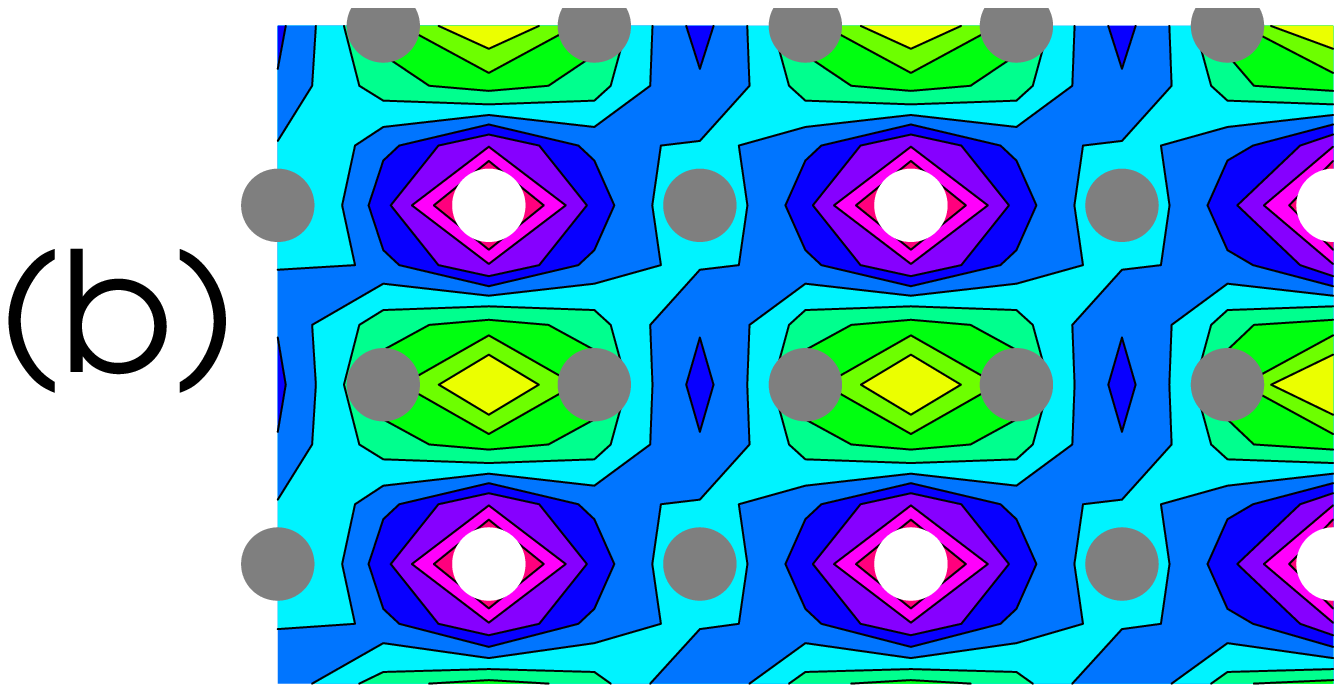}
  \caption{Na Coulomb potential in Co planes at $x={1\over2}$ with divacancies
aligned along {\bf c} axis (a) and displaced by {\bf b} (b). The white (black)
circles represent Co$^{\approx 3+}$ (Co$^{\approx 4+}$) ions. 
The gray circles represent conducting Co ions at intermediate valencies.}
\end{figure}


%
\end{document}